Research Article

# Vibe-FDTR: An agent-oriented framework for reproducible frequency-domain thermoreflectance data analysis

Fuwei Yang[a,b,c,†], Weiheng Li[a,d,†], Bai Song[a,d,*]

[a] National Key Laboratory of Advanced Micro and Nano Manufacture Technology, Peking University, Beijing, 100871, China
[b] Center for Nano and Micro Mechanics, Tsinghua University, Beijing, 100084, China
[c] Department of Engineering Mechanics, Tsinghua University, Beijing, 100084, China
[d] School of Mechanics and Engineering Science, Peking University, Beijing, 100871, China

ARTICLE INFO

ABSTRACT

Frequency-domain thermoreflectance (FDTR) is a laser pump-probe technique widely used to measure thermal properties at the micro- and nanoscale; however, it relies on a complex data analysis procedure that demands substantial domain expertise and is susceptible to subtle human errors. Here, we present Vibe-FDTR, an agent-oriented framework that enables large language model (LLM) agents to perform reliable and reproducible FDTR analyses directly from natural language requests. This framework couples a configuration-driven FDTR code package, which enforces physical and parametric consistency, with procedural agent skills that translate user intentions into organized and verifiable analysis steps. We evaluate Vibe-FDTR using a controlled benchmark with two levels: synthetic single-step tasks and real-data multi-step tasks based on measurements of gold-coated graphite samples. Across the two levels, agents using Vibe-FDTR achieve success rates of 100% and 98.9%, respectively. In sharp contrast, ablating skills (Code-agent) reduces performance to 91.4% and 36.7%, which drops further to 38.6% and 0% when the domain package is also omitted (Agent-only). Beyond success rates, Vibe-FDTR also reduces computational cost by 87.7% relative to the Code-agent variant and cuts execution time by more than 60%. Finally, an optional expert mode supports experimental planning via autonomous sensitivity and uncertainty evaluations, and formulates physically grounded recommendations for underspecified tasks. These results demonstrate that encapsulating domain code and expert knowledge into agent skills offers a promising route toward low-barrier, autonomous, and trustworthy thermal metrology.

## 1. Introduction

Accurate measurement of thermal properties at the micro- and nanoscale is essential for understanding heat transport and guiding the thermal design of modern electronic and energy systems.[1–5] To probe thermal transport in samples of diverse sizes, structures, and dimensions, a variety of experimental approaches have been developed, including the 3ω method,[6,7] scanning thermal microscopy (SThM),[8,9] thermal bridge,[10–12] opto-thermal Raman spectroscopy,[13,14] and all sorts of laser pump-probe techniques.[15,16] Among these approaches, time-domain thermoreflectance (TDTR) and frequency-domain thermoreflectance (FDTR) are widely recognized for their reliability, versatility, and relative ease of use.[17,18] By using a modulated pump laser to generate a localized thermal response and a probe laser to monitor the corresponding surface thermoreflectance variation, TDTR and FDTR can resolve heat transport over fast time scales and small length scales with high sensitivity, enabling the precise determination of thermal conductivity, specific heat, and interfacial thermal conductance. Furthermore, these techniques are inherently non-contact and eliminate the need for integrating electrical heaters or thermometers onto the sample.[7,11] In light of these advantages, TDTR and FDTR have been extensively employed to characterize both bulk materials and thin films, with thermal conductivity values spanning five orders of magnitude from ultralow (~0.01 $\mathrm{W\,m^{-1}\,K^{-1}}$) to ultrahigh (>1000 $\mathrm{W\,m^{-1}\,K^{-1}}$).[19–24] In addition, TDTR and FDTR have also been expanded to buried interfaces and microfabricated structures, and combined with in

* Corresponding author at: School of Mechanics and Engineering Science, Peking University, Beijing 100871, China.
*E-mail address:* songbai@pku.edu.cn (B. Song).
† These authors contributed equally to this work.

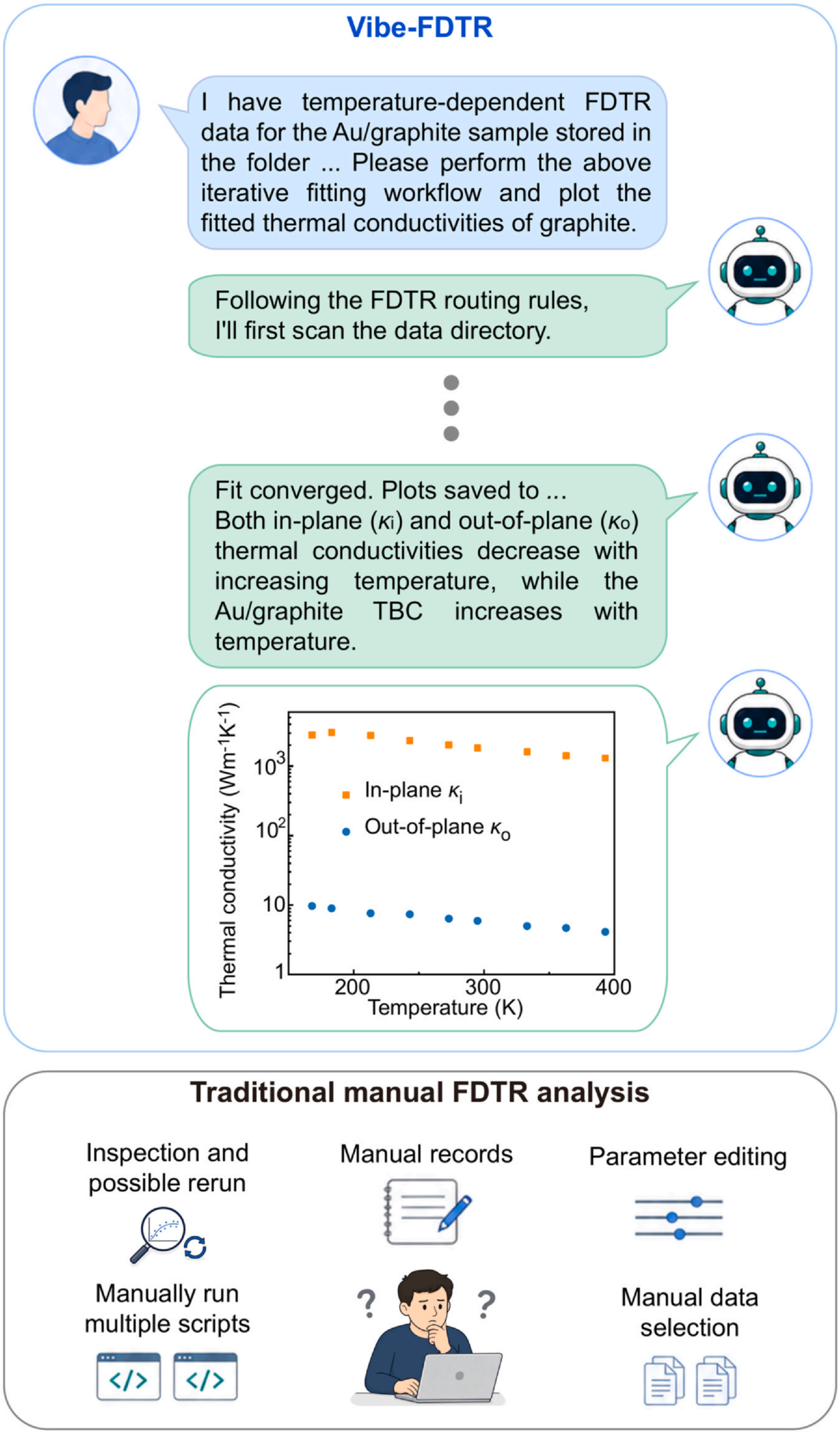

**Fig. 1.** Schematic illustration of Vibe-FDTR compared with the traditional manual FDTR analysis workflow.

situ experiments involving external fields or mechanical manipulations.[25–29]

Despite the widespread application of TDTR and FDTR, a complex and highly demanding data-processing workflow is required to accurately extract the target thermal properties. Unlike the thermal bridge method where thermal conductivity can be readily derived via Fourier's law from the applied heat flux and measured temperature difference,[7,11] thermoreflectance measurements rely on an indirect model fitting procedure.[30] This nonlinear inverse problem is inherently limited by parameter sensitivities and correlations, and therefore a single signal or fitting window often provides insufficient identifiability for extracting multiple thermal properties. To obtain reliable results, complementary information from different signals may be required, such as those measured with different laser spot sizes or pump-probe offset distances.[28,31,32] In practice, a complete thermoreflectance data analysis is rarely a single fitting operation, but rather a sequence of coupled decisions involving data organization, sensitivity evaluation, fitting-parameter selection, and uncertainty assessment, as illustrated in Fig. 1. Such a workflow is tedious and inefficient when performed manually, especially for large datasets or iterative analyses, and is susceptible to subtle human errors. Minor mistakes may still lead to apparently successful code execution, but ultimately produce physically invalid results.

Very recently, the synergistic development of various large language models (LLMs) and harnesses has enabled artificial intelligence (AI) to not only serve as isolated prediction and acceleration tools, but also act as agents that proactively execute workflows. Leading open-source LLMs such as DeepSeek-V4 [33] have focused increasingly on agentic capabilities. By enabling AI to interpret user intent and translate it into concrete actions—such as configuration writing, tool calling, and data management—this progress has the potential to significantly improve the efficiency of scientific research. In fact, such a paradigm shift has already been preliminarily demonstrated in a variety of settings, including chemistry agents equipped with domain tools and LLM-guided experimental planning,[34] closed-loop materials synthesis,[35] automated atomic force microscopy,[36] and AI-driven scanning probe microscopy.[37] Of central importance is the establishment of agentic frameworks that use procedural guidance to translate high-level scientific intent into traceable tool invocations and verifiable outputs. In thermoreflectance metrology, however, previous efforts with AI have mainly focused on solving data-driven inverse problems, including neural networks for inferring thermal parameters from TDTR signals,[38] reconstructing depth-dependent thermal conductivity from FDTR responses,[39] and evaluating interfacial bond quality from FDTR phase maps.[40] To the best of our knowledge, no agent-oriented framework has been reported to reliably and reproducibly perform TDTR/FDTR analysis with natural language requests, which can minimize the need for expert knowledge and largely reduce operational burden associated with thermoreflectance metrology.

Here, we take FDTR as an example and present Vibe-FDTR, an agent-oriented framework inspired by recent vibe-coding practices that translate natural language requests into executable code through LLM-based agents.[41–43] Vibe-FDTR adapts this idea to thermal metrology by combining domain-specific procedural skills and a configuration-driven FDTR code package. The framework is designed primarily to reliably execute well-specified post-measurement tasks, as demonstrated in Fig. 1 with a representative natural language input; while an optional expert mode adds an additional skill layer for FDTR experimental planning and for analysing underspecified problems. To quantitatively evaluate agent performance, we employ a controlled two-level benchmark comprising both single-operation tasks on synthetic data and multi-step tasks on real data obtained from gold-coated graphite samples. Through repeated runs, the high success rates, low computational cost, and short execution time of Vibe-FDTR are systematically quantified and highlighted. For comparison, two reference scenarios are considered, one with the skills ablated (Code-agent) and the other with the domain package further omitted (Agent-only). Finally, we demonstrate the expert mode as an intelligent copilot to autonomously perform sensitivity and uncertainty analyses, and provide physically grounded recommendations regarding a series of partially specified tasks.

## 2. Principles of FDTR measurement and data analysis

### 2.1. FDTR experimental setup

We first briefly introduce the FDTR platform and basic measurement principle using our experimental setup as a representative example.[23,24,28] As shown in Fig. 2(a), the platform employs a two-laser pump-probe configuration, in which a 405 nm continuous-wave (CW) pump laser is intensity-modulated by the reference output of a lock-in amplifier and focused onto the metal-coated sample surface to generate a periodic temperature field. A 532 nm CW probe laser is aligned to the heated region and monitors the resulting temperature oscillation through the temperature-dependent reflectance of the metal transducer (typically gold). Both the pump and probe beams feature a Gaussian cross-sectional intensity profile. The reflected probe beam is converted into an electric current via a balanced photodetector to improve the signal-to-noise ratio. Subsequently, the lock-in amplifier is used to resolve the amplitude and phase signals relative to the pump modulation. In the most commonly used frequency-sweep ($f$-sweep) mode, the modulation frequency is varied to obtain the corresponding thermal response, while beam-offset measurements scan the lateral pump-probe separation at fixed $f$, which is superior for probing in-plane heat spreading and for characterizing the spot size. The schematics of the two measurement modes are demonstrated in the insets of Fig. 2b and c. Sample positioning is achieved using a piezoelectric stage, and temperature-dependent measurements are performed by integrating a heating/cooling stage (INSTEC HCP421V in this work) on top of it.

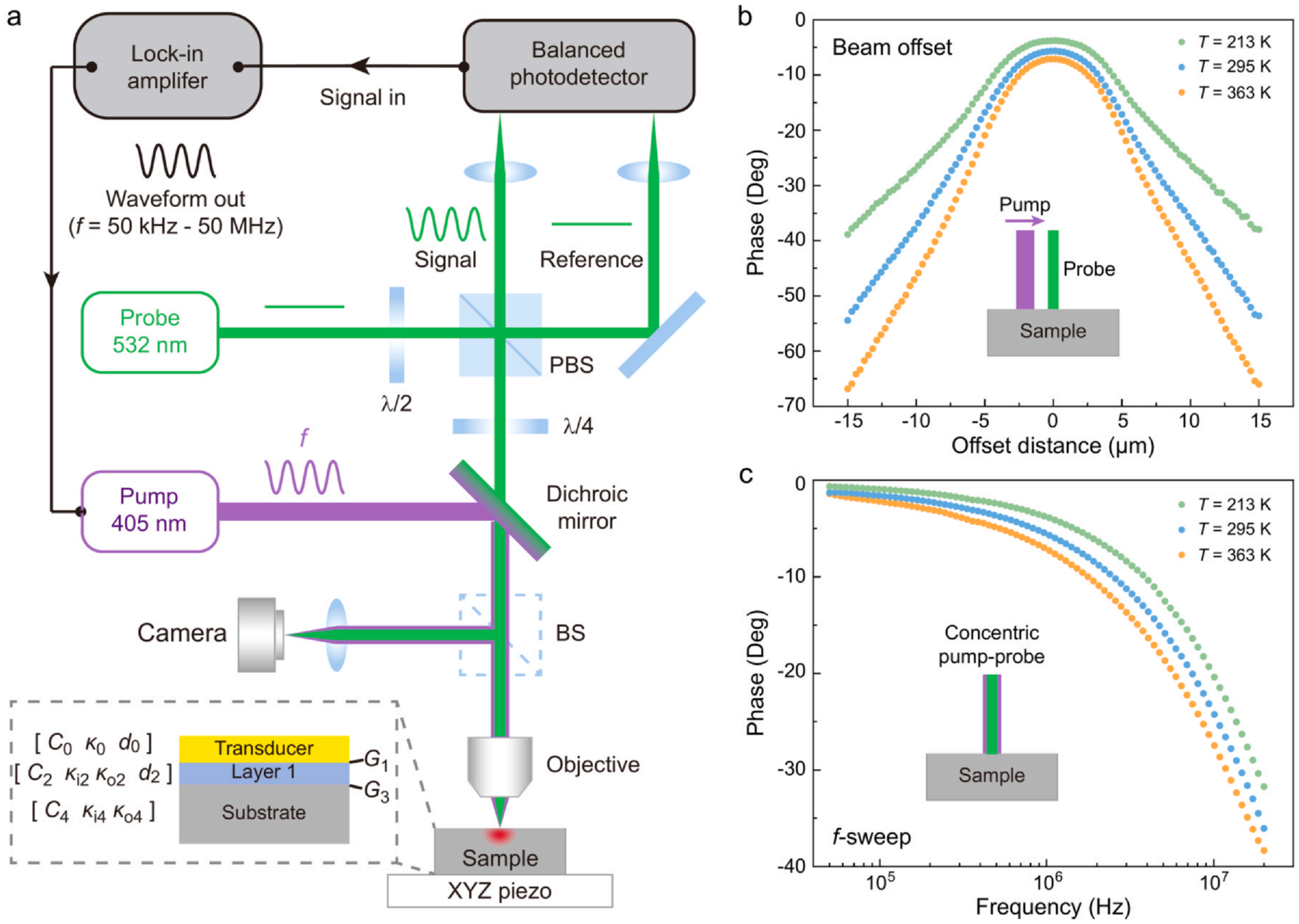

**Fig. 2.** Basic principles of the FDTR technique. (a) Schematic of FDTR experimental setup. (b) Representative beam-offset phase signals at $f = 1$ MHz measured on an Au/graphite sample at three different temperatures. (c) Representative $f$-sweep signals measured on the same sample. The insets show the schematics of the two measurement modes.

### 2.2. Thermal model for fitting

After obtaining the amplitude and phase signals, thermal properties of the sample are extracted by comparing the measured response with a multilayer Fourier heat diffusion model.[21] With an appropriately small pump power, the thermoreflectance signal of the metal transducer remains linear with respect to the surface temperature oscillation. The absorbed pump power is commonly approximated as a modulated Gaussian heat source at the transducer surface. The current FDTR code package supports isotropic and transversely isotropic thermal models, and more complex models can be readily incorporated in future implementations when needed.

For a transversely isotropic layer, the heat conduction equation is given by

$$\kappa_{\mathrm{i}}\left(\frac{\partial^2 T}{\partial r^2}+\frac{1}{r}\frac{\partial T}{\partial r}\right)+\kappa_{\mathrm{o}}\frac{\partial^2 T}{\partial z^2}=C\frac{\partial T}{\partial t}, \tag{1}$$

where $\kappa_{\mathrm{i}}$ and $\kappa_{\mathrm{o}}$ are the in-plane and out-of-plane thermal conductivities, respectively, and $C$ is the volumetric heat capacity. After transformation into the frequency domain and Hankel space, the temperature and heat flux at the two surfaces of each layer can be related through a transfer matrix as

$$\begin{pmatrix}\Theta_{\mathrm{b}}\\ Q_{\mathrm{b}}\end{pmatrix}=\boldsymbol{M}\begin{pmatrix}\Theta_{\mathrm{t}}\\ Q_{\mathrm{t}}\end{pmatrix}, \tag{2}$$

where $\Theta$ and $Q$ denote the transformed temperature and heat flux, respectively; and the subscripts t and b refer to the top and bottom surfaces of the layer, respectively. For a homogeneous layer of thickness $d$, the matrix is

$$\boldsymbol{M}=\begin{bmatrix}\cosh(\eta d) & -\frac{1}{\kappa_{\mathrm{o}}\eta}\sinh(\eta d)\\ -\kappa_{\mathrm{o}}\eta\sinh(\eta d) & \cosh(\eta d)\end{bmatrix}, \tag{3}$$

$$\eta^2=\frac{\kappa_{\mathrm{i}}\beta^2+2\pi i f C}{\kappa_{\mathrm{o}}}, \tag{4}$$

where $\beta$ is the Hankel transform variable and $f$ is the modulation frequency. An interface with thermal conductance $G$ is described by

$$\boldsymbol{M}_{\mathrm{int}}=\begin{bmatrix}1 & -1/G\\ 0 & 1\end{bmatrix}. \tag{5}$$

The complete multilayer response is obtained by multiplying the transfer matrices of all layers and interfaces. Denoting the total transfer matrix as

$$\boldsymbol{M}_{\mathrm{tot}}=\begin{bmatrix}A_M & B_M\\ C_M & D_M\end{bmatrix}, \tag{6}$$

and applying the bottom boundary condition, the complex surface temperature response can be calculated and averaged over the probe-beam intensity profile as

$$H(f,x_0)=\iint\frac{2}{\pi w_1^2}\exp\left[-\frac{2((x-x_0)^2+y^2)}{w_1^2}\right]\times\left[\frac{P_0}{2\pi}\int_0^\infty\beta J_0(\beta\sqrt{x^2+y^2})\left(-\frac{D_M}{C_M}\right)\exp\left(-\frac{\beta^2 w_0^2}{8}\right)d\beta\right]\mathrm{d}x\mathrm{d}y. \tag{7}$$

Here, $P_0$ is the absorbed pump power, $w_0$ and $w_1$ are the pump and probe beam radii, respectively, $J_0$ is the zeroth-order Bessel function, and $x_0$ is the lateral pump-probe offset. The modeled complex response $H(f,x_0)$ provides the calculated FDTR amplitude and phase. Thermal properties are then obtained by adjusting selected model parameters, such as thermal conductivity, heat capacity, or interfacial thermal conductance, until the modeled response best matches the measured frequency-sweep and/or beam-offset data. Independently calibrated quantities, such as transducer thickness and laser spot size, are usually fixed during fitting.

### 2.3. Sensitivity and uncertainty analysis

Because FDTR extracts thermal properties through nonlinear fitting, sensitivity analysis is employed to assess whether a given set of parameters can be reliably fitted from the selected signal and data range. For an observable $y$, such as the measured amplitude or phase, the normalized sensitivity to a parameter $x$ is defined as [22]

$$S_x=\frac{\partial y}{\partial x}\Big/\frac{y}{x}. \tag{8}$$

This quantity describes the relative change in the calculated signal caused by a small variation in the parameter of interest.

The uncertainty of the fitted parameters can be estimated by Monte Carlo simulations or a full-error propagation formula.[44] Here, we adopt the latter method. Briefly, the fitting process can be written as the minimization of the residual between the measured and modeled FDTR responses:

$$R=\sum_{i=1}^{M_{\mathrm{p}}}[y_i-f(\alpha_i,X_U,X_C)]^2. \tag{9}$$

Here, $\alpha_i$ is the scanned variable such as the modulation frequency or pump–probe offset, $y_i$ is the measured signal at $\alpha_i$, $X_U$ denotes the fitted unknown parameters, and $X_C$ denotes the fixed input parameters. Let $F$ and $Y$ be the $M_{\mathrm{p}}\times 1$ vectors formed by $f(\alpha_i,X_U,X_C)$ and $y_i$, respectively. The Jacobian matrices with respect to the fitted and fixed parameters are

$$J_U=\frac{\partial F}{\partial X_U},\quad J_C=\frac{\partial F}{\partial X_C}. \tag{10}$$

The variance–covariance matrix of the fitted parameters is then given by

$$Var(X_U)=(J_U^*J_U)^{-1}J_U^*[Var(Y)+J_C Var(X_C)J_C^*]J_U(J_U^*J_U)^{-1}, \tag{11}$$

where the superscript $*$ denotes the complex conjugate transpose. Here, $\mathrm{Var}(Y)$ represents the variance of the experimental signal, and $\mathrm{Var}(X_C)$ represents the uncertainties of the fixed input parameters. The contribution from $\mathrm{Var}(Y)$ is often negligible[23,24,28] and therefore ignored in the current code package. The square roots of the diagonal elements of $\mathrm{Var}(X_U)$ yield the uncertainties of the fitted parameters. We note that such uncertainty analysis is local in the parameter space, and does not capture optimization-induced errors associated with convergence to a local minimum.

## 3. Vibe-FDTR framework

We first introduce the architectural design of the Vibe-FDTR framework. As shown in Fig. 3, the core of Vibe-FDTR consists of a code layer and a procedural guidance layer, with the former enforcing physical and parametric consistency, while the latter governing how user requests are translated into organized and verifiable analyses. Together, the two layers make agent executions predictable and auditable. Furthermore, an optional expert layer extends this core architecture when users explicitly request support for experimental design, dealing with underspecified problems, or failure diagnosis. In such cases, the expert skill provides additional structured scientific guidance for AI agents based on predefined expert experience.

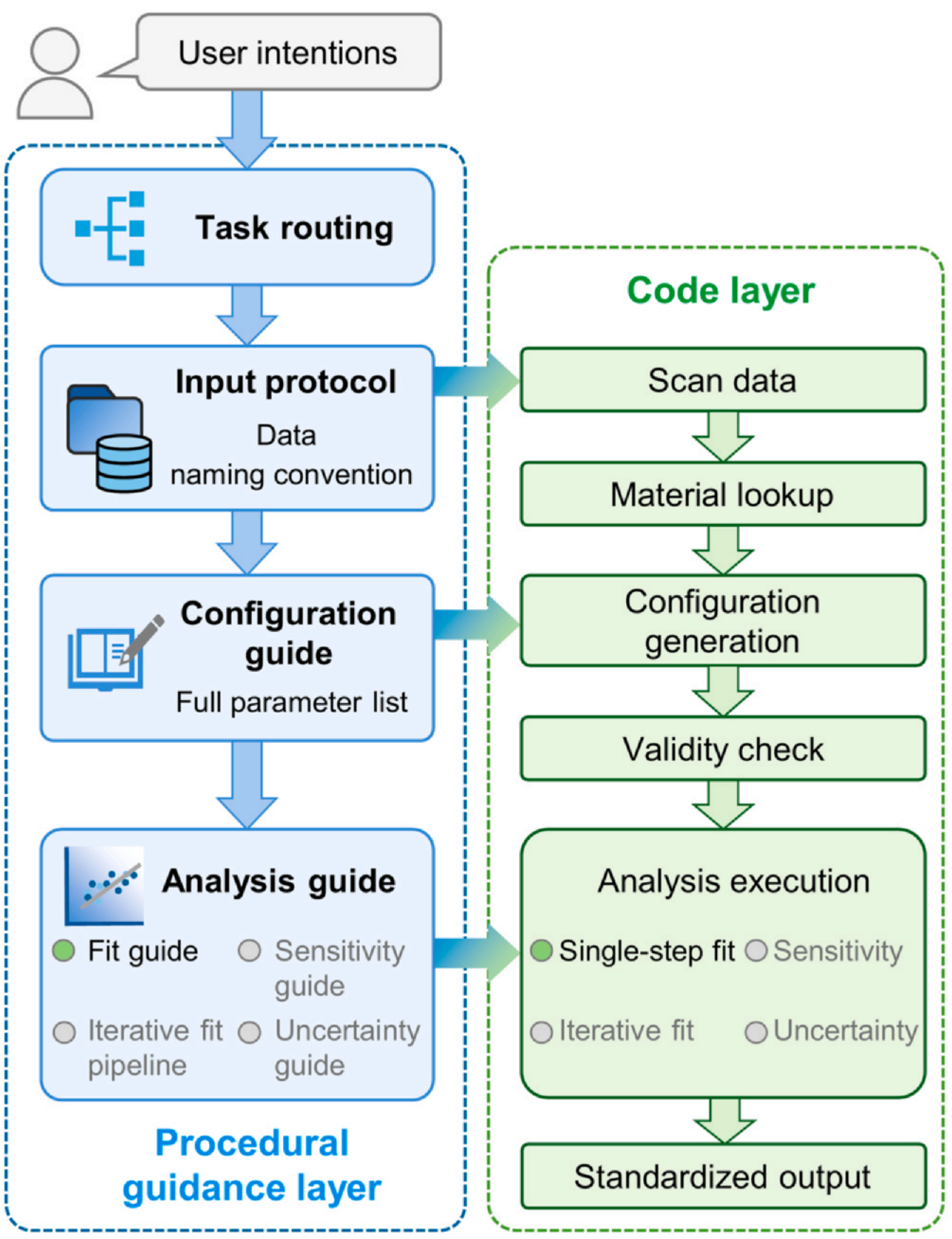

**Fig. 3.** Core architecture of Vibe-FDTR which is composed of a procedural guidance layer and a code layer.

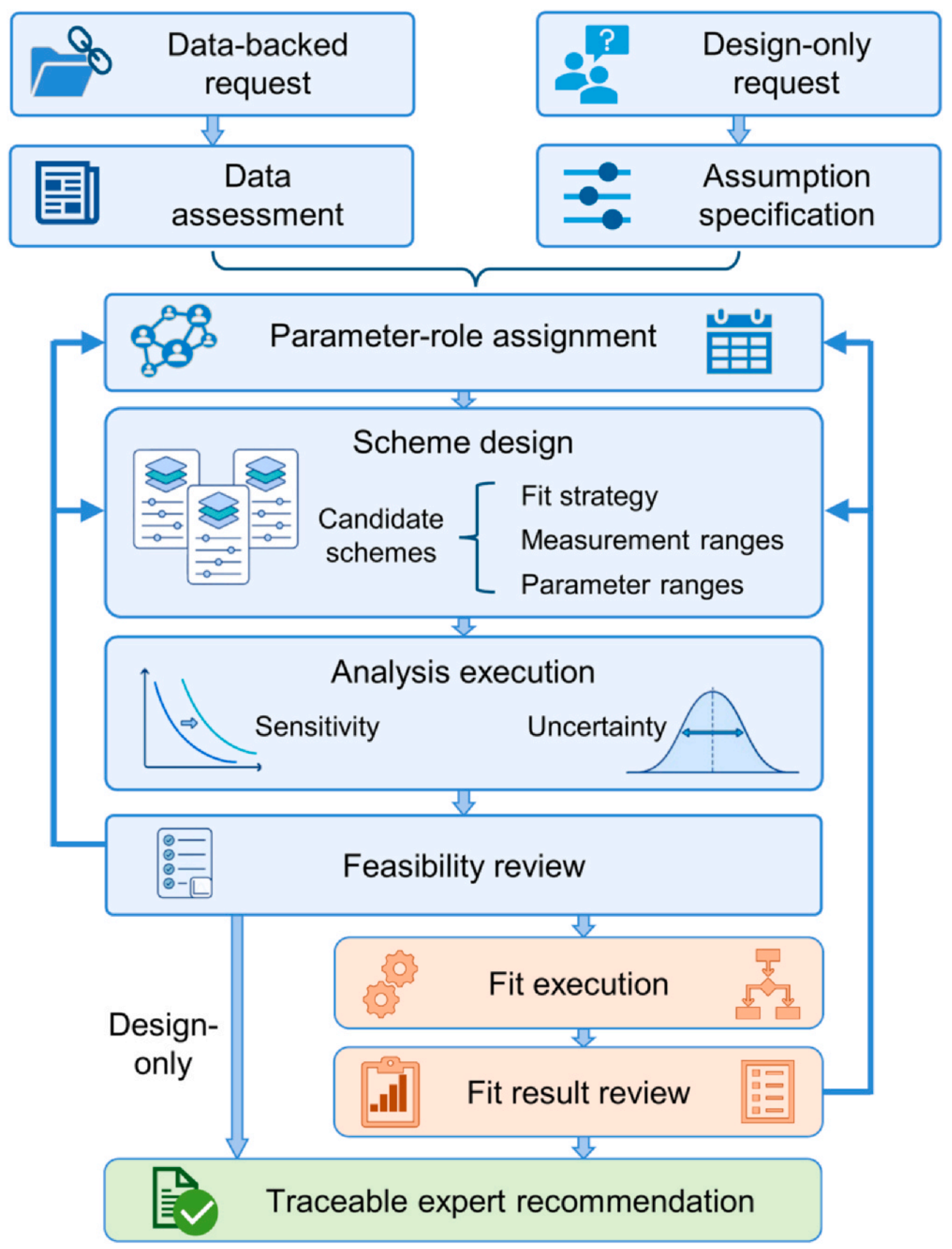

**Fig. 4.** Optional expert mode workflow of Vibe-FDTR.

### 3.1. Code layer

The code layer provides the execution foundation of Vibe-FDTR. Conventional FDTR analysis often relies on separate scripts for different processing and fitting operations, requiring manual transfer of parameters and intermediate results. Vibe-FDTR instead integrates data discovery, material-property lookup, model configuration, numerical analysis, and standardized output within a unified package (Fig. 3). All modules share consistent parameter definitions, unit conventions, and data structures, allowing successive analysis steps to operate on a common basis. A clear interface also separates agent-directed workflow construction from the implementation of the thermal model and numerical solvers, so that the scientific validity of the analysis is anchored in the code layer, which has been validated across multiple material systems in our previous studies.[23,24,28,45–47]

A central design choice of the code layer is to use configuration files as persistent, machine-readable specifications. These files preserve the model definition and analysis settings so that each calculation can be inspected, reproduced, or modified without reassembling information from dispersed files or relying on manual records. Multi-step analysis is described by a separate pipeline specification that defines the execution order and transfers fitted quantities between successive steps. Before numerical execution, the code layer validates the configuration files and rejects incomplete, incompatible, or physically invalid inputs. This code-level validation provides the hard constraints of Vibe-FDTR, setting the stage for the procedural rules encoded in the guidance layer.

### 3.2. Procedural guidance layer

The procedural guidance layer translates user requests into executable FDTR operations through agent skills. Even when a user provides a clear analysis request, the agent still needs to map that request to the correct code entry point, construct the required configuration, select the appropriate analysis module, and pass intermediate files between steps. The role of this layer is therefore to provide task-level guidance for code use, so that the agent does not need to infer the analysis logic via the underlying source code.

Vibe-FDTR uses a progressive-disclosure structure to limit the information exposed to the agent at each stage. The main skill defines the overall workflow, parameter-naming conventions, supported analysis functions, and standard execution order, including frequency-sweep fitting, beam-offset fitting, spot-size fitting, sensitivity analysis, uncertainty propagation, and iterative fitting. More specialized skills are loaded only when required. Configuration generation is handled by a dedicated skill that converts the sample structure, material properties, fixed inputs, fitting targets, parameter bounds, signal channels, and data ranges into a machine-readable configuration file. Sensitivity and uncertainty analyses are handled by separate skills that extend an existing fitting configuration with the additional information needed for reliability evaluation. For multi-step analysis involving parameter transfer, an iterative-pipeline skill specifies the fitting sequence, update rules, convergence criteria, and required artifacts.

### 3.3. Expert layer

Built on top of the code layer and the procedural guidance layer, the expert layer is implemented as an optional upper-level skill which is invoked only by explicit user requests. Routine tasks therefore remain direct, while more ambiguous requests can be examined within a broader FDTR reasoning procedure. With this skill, the agent can better interpret the overall FDTR data-processing workflow and incorporate relevant prior experience, therefore facilitating less experienced users with complicated tasks.

As shown in Fig. 4, the expert skill first distinguishes between design-only and data-backed requests. For design-only tasks, missing physical quantities are assigned explicit physically reasonable assumptions, followed by parameter-role assignment to separate fitting parameters from constrained inputs. The skill then develops candidate analysis schemes and evaluates them through sensitivity and uncertainty calculations performed by the underlying task skills. The results are reviewed, and the scheme is revised with additional calculations when needed before a structured recommendation is produced.

For data-backed tasks, the workflow begins with data-quality screening to identify corrupted points, spikes, or inconsistent repeats. It then applies the same pre-fit design process to select a justifiable fitting scheme. After execution, the fitted results are reviewed from several aspects including residual quality and physical plausibility, with unsatisfactory results triggering the revision of the assumptions or fitting schemes. Throughout the workflow, each decision stage is informed by expert experience which is extracted through AI-mediated elicitation of human FDTR experience from representative cases.

## 4. Benchmark design

### 4.1. Benchmark tasks

To quantitatively evaluate Vibe-FDTR for the reliable execution of well-defined FDTR data analysis tasks, we constructed a two-level benchmark with seven Level-1 (L1) tasks and nine Level-2 (L2) tasks. These tasks cover commonly used analysis methods, including both $f$-sweep and beam-offset fitting, as well as sensitivity and uncertainty calculations. The L1 tasks are simple single-step operations based on synthetic FDTR data generated from the forward model described in Section 2, which are intended to test the reliability of an AI agent in the most straight-forward scenario. In terms of materials selection, we consider representative isotropic and anisotropic glasses and crystals, including fused silica ($SiO_2$), silicon (Si), graphite, and hexagonal boron nitride (hBN). For L2 tasks, we employ real FDTR experimental data from a gold-coated graphite (Au/graphite) sample. More complex workflows are designed, involving combined data preprocessing, analysis, multiple fitting steps and batch processing of different data groups. The L2 tasks closely reflect the end-to-end workflow typically followed by researchers in real-world settings. The key inputs for all these tasks are summarized in Table 1, with the ground truth independently determined via manual inspection.

The Au/graphite specimen was prepared from commercially available highly oriented pyrolytic graphite (HOPG, ZYA grade, NT-MDT). Before metal deposition, the HOPG was mechanically cleaved to expose a fresh basal-plane surface. The graphite specimen and a fused-silica reference sample were mounted on the same silicon carrier wafer and coated in a single electron-beam evaporation process. The metal transducer consisted of a Ti adhesion layer and an Au layer. The Au thickness was obtained by measuring patterned step edges on the fused-silica reference using atomic force microscopy (AFM, Cypher ES, Asylum Research). The temperature-dependent electrical conductivity of the Au film was measured using a standard four-probe method, and its thermal conductivity was calculated from the Wiedemann-Franz law. In Fig. 2b and c, we demonstrate the measured beam-offset and $f$-sweep phase signals at three representative temperatures.

Beyond the well-specified L1 and L2 tasks, a set of partially specified requests without quantitative ground truth further challenges the optional expert mode to autonomously plan sensitivity and uncertainty analyses and to condense the resulting evidence into physically grounded recommendations. These underspecified expert-mode (E) tasks are summarized in Table 2. Spanning from routine fitting requests to open-ended experimental design, this layered benchmark allows a comprehensive examination of where the Vibe-FDTR framework excels and where its current limits lie.

### 4.2. Evaluation protocol

We build a containerized environment to execute and record automated FDTR analyses using the Vibe-FDTR framework. To isolate the respective contributions of the domain code and the procedural skills, each task is additionally executed under two reference configurations, namely a Code-agent variant with the skill documents ablated and an Agent-only setting where the domain package is also withheld.

As shown in Fig. 5, this environment is based on OpenCode,[48] an open-source harness which we connect to the DeepSeek-V4-Pro API. Each run takes place in an isolated Docker container. At the beginning of a test, the evaluation program populates the container with the required inputs and injects the task prompt; the agent subsequently carries out the request without human intervention. For the Code-agent and Agent-only settings, where no skill documents are available, we additionally inform the agent of the runtime details. Upon completion, the output files, total wall-clock time, and API call cost are collected for assessment.

All tests are conducted on a dual-socket EPYC 9654 workstation with 192 cores and ample memory. Fifty containers run in parallel, compressing the full suite into approximately one hour and thereby minimizing temporal fluctuations in API service performance. Monitoring of the computational resources confirms that CPU and

**Table 1**
Overview of L1 and L2 benchmark tasks. L1 denotes single-operation tests on synthetic datasets, and L2 denotes integrated workflows on measured Au/graphite data.

| ID | Operation | Sample | Input constraint | Main targets |
|---|---|---|---|---|
| L1-F01 | Frequency fit | Au/fused silica | Fixed $\kappa_{SiO_2}$ and spot size | $\kappa_{Au}$, $G$ |
| L1-F02 | Frequency fit | Au/Si | Fixed $\kappa_{Au}$ and spot size | $d_{Au}$, $\kappa_{Si}$, $G$ |
| L1-F03 | Frequency fit | Au/graphite | Fixed $\kappa_{i,Gr}$ and spot size | $\kappa_{o,Gr}$, $G$ |
| L1-F04 | Frequency fit | Au/hBN/Si | Fixed $\kappa_{o,hBN}$ and hBN/Si $G$ | $\kappa_{i,hBN}$, Au/hBN $G$ |
| L1-O01 | Beam-offset fit | Au/graphite | Fixed $\kappa_{o,Gr}$ and modulation frequency | $\kappa_{i,Gr}$, $G$ |
| L1-S01 | Sensitivity analysis | Au/fused silica | Specified parameters and spot sizes | Sensitivity to $\kappa_{SiO_2}$, $G$, and $d_{Au}$ |
| L1-U01 | Uncertainty propagation | Au/graphite/Si | Prescribed fixed-parameter uncertainties | Uncertainties of $\kappa_{i,Gr}$ and $\kappa_{o,Gr}$ |
| L2-D01 | Averaging + frequency fit | Au/graphite | RT; one repeated point excluded | $\kappa_{o,Gr}$, $G$ |
| L2-D02 | Averaging + beam-offset fit | Au/graphite | 60 °C; selected offset window | $\kappa_{i,Gr}$, $G$ |
| L2-C01 | Sensitivity + frequency fit | Au/graphite | RT; selected frequency range | Sensitivity, $\kappa_{o,Gr}$, $G$ |
| L2-C02 | Beam-offset fit + uncertainty | Au/graphite | RT; selected offset window | $\kappa_{i,Gr}$, $G$, uncertainty |
| L2-C03 | Sensitivity + fit + uncertainty | Au/graphite | RT | Sensitivity, $\kappa_{o,Gr}$, $G$, uncertainty |
| L2-I01 | Spotfit + frequency fit | Au/graphite | RT; highest-frequency amplitude | Spot size, $\kappa_{o,Gr}$, $G$ |
| L2-I02 | Iterative fit + uncertainty | Au/graphite | RT; spotfit, offset fit, and frequency fit | Spot size, $\kappa_{i,Gr}$, $\kappa_{o,Gr}$, $G$, uncertainty |
| L2-B01 | Temperature-batch fitting | Au/graphite | Temperature-dependent data | $\kappa_{i,Gr}(T)$, $G(T)$ |
| L2-B02 | Temperature-batch iteration | Au/graphite | RT spotfit followed by temperature-dependent iteration | Spot size, $\kappa_{i,Gr}(T)$, $\kappa_{o,Gr}(T)$, $G(T)$ |

**Table 2**
Summary of expert-mode benchmark tasks and the final outputs by the agent using Vibe-FDTR. Here, only key information is selected for display convenience.

| ID | User intention | Vibe-FDTR recommendations |
|---|---|---|
| E-1 | Design FDTR measurement for Au/diamond. Target: $\kappa_{\text{diamond}}$. | Assume a nominal 3 µm spot size and heat capacities from the database. Recommend $f$-sweep fitting over 5 kHz-20 MHz for $\kappa_{\text{diamond}}$ and $G_{\text{Au/diamond}}$. May use high-frequency offset amplitude for spot calibration and offset phase fit as a cross-check. Justify both fitted quantities with about 8% uncertainty. |
| E-2 | Design FDTR measurement for Au/graphite. Targets: $\kappa_{\text{i,Gr}}$ and $\kappa_{\text{o,Gr}}$. | Assume the spot size is unknown and use heat capacities from the database. Recommend a sequential fitting workflow with 50 MHz offset amplitude for spot fitting, 5 MHz offset phase for $\kappa_{\text{i,Gr}}$, and 100 kHz-50 MHz $f$-sweep phase for $\kappa_{\text{o,Gr}}$ and $G_{\text{Au/Gr}}$. Justify by sensitivity and uncertainty analysis. |
| E-3 | Design FDTR measurement for Au/graphene/Si. Targets: $\kappa_{\text{i,Gr}}$ and $\kappa_{\text{o,Gr}}$. | Assume graphene heat capacity from graphite-based data and fix $G_{\text{Gr/Si}}$ at $5 \times 10^7$ W m$^{-2}$ K$^{-1}$. Recommend beam-offset fitting for $\kappa_{\text{i,Gr}}$ and $f$-sweep fitting for $G_{\text{Au/Gr}}$. Reject simultaneous fitting of $\kappa_{\text{o,Gr}}$ and $G_{\text{Au/Gr}}$ because film and interfacial thermal conductance are degenerate. |
| E-4 | Recommend Au thickness for FDTR measurement of sapphire. Consider different possible $G$. Target: $\kappa_{\text{sapphire}}$. | Assume a 3 µm spot size and heat capacities from the database. Recommend 70–80 nm thick Au and fit $\kappa_{\text{sapphire}}$ and $G$ by $f$-sweep fitting over 50 kHz–10 MHz. Justify by batch sensitivity runs with different $d_{\text{Au}}$ and $G$, and uncertainties at 80 nm (8.4% and 6.2% for $\kappa_{\text{sapphire}}$ and $G$, respectively). |
| E-5 | Recommend spot size and frequency range for $f$-sweep measurement on Au/graphite. Targets: $\kappa_{\text{i,Gr}}$ and $\kappa_{\text{o,Gr}}$. | Assume graphite heat capacity from the database. Recommend a 5–8 µm spot and a 200 kHz–20 MHz $f$-sweep window. Fit $\kappa_{\text{i,Gr}}$, $\kappa_{\text{o,Gr}}$, and $G_{\text{Au/Gr}}$ simultaneously. Justify by sensitivity analysis and uncertainties at different spot sizes. Warn about the local minima risk for three-parameter fitting. |
| E-6 | Recommend the reliable fitting range for $G$ of Au/fused silica. Target: $G$ from $10^6$ to $10^9$ W m$^{-2}$ K$^{-1}$. | Assume a 3 µm spot size and thermal properties from the database. Recommend reliable fitting for $G \le 3 \times 10^7$ W m$^{-2}$ K$^{-1}$, with an optimum near $10^7$ W m$^{-2}$ K$^{-1}$; reject $G \ge 10^8$ W m$^{-2}$ K$^{-1}$ as poorly identifiable from $f$-sweep phase. Justify with sensitivity and uncertainty results for $G$ from $10^6$ to $10^9$ W m$^{-2}$ K$^{-1}$ |
| E-7 | Explore separable fitting conditions for Au/thin film/SiC. Consider a typical thin-film heat capacity. Targets: film $\kappa$ and the two $G$. | Assume an 80 nm Au transducer, thin-film heat capacity of 2.5 MJ m$^{-3}$ K$^{-1}$, $G_{\text{Au/film}}$ = 200 MW m$^{-2}$ K$^{-1}$, and a 3 µm spot size. Identify poor separability of film $\kappa$ and $G_{\text{film/SiC}}$ in a single measurement. Use $d_{\text{film}}G/\kappa_{\text{film}}$ to decide whether to fit film $\kappa$ or interface $G$ while fixing the other. Justify by sensitivity and uncertainty results for different film thickness, film $\kappa$, and $G_{\text{film/SiC}}$. |
| E-8 | Explore identifiable co-fitting sets for Au/fused silica $f$-sweep phase data. Targets: properties of Au and $SiO_2$, spot size, and $G$. | Assume about 48 nm Au, 3 µm spot size, $G_{\text{Au/SiO}_2} \approx 50$ MW m$^{-2}$ K$^{-1}$, and two-parameter fitting as the practical limit. Recommend viable parameter pairs and reject strongly correlated Au-property and substrate-property sets by sensitivity and uncertainty results for different co-fitting sets involving $d_{\text{Au}}$, $\kappa_{\text{Au}}$, $C_{\text{Au}}$, spot size, $\kappa_{\text{SiO}_2}$, $C_{\text{SiO}_2}$, and $G$. |
| E-9 | Execute $f$-sweep fitting for Au/graphite using the given dataset. Targets: $\kappa_{\text{o,Gr}}$ and $G$. (Bad data with phase spike present, which is the same dataset used in L2-D01) | Detect and exclude the point11 data file with phase spikes and fit 10 clean $f$-sweep repeats. Report $\kappa_{\text{o,Gr}}$ = 5.9 W m$^{-1}$ K$^{-1}$ and $G = 9.4 \times 10^7$ W m$^{-2}$ K$^{-1}$. |
| E-10 | Choose proper fitting methods for Au/hBN/Si using the given data. Targets: $\kappa_{\text{i,hBN}}$ and $\kappa_{\text{o,hBN}}$. | Choose iterfit with offset phase for $\kappa_{\text{i,hBN}}$ and $f$-sweep phase for $\kappa_{\text{o,hBN}}$, $G_{\text{Au/hBN}}$, and $G_{\text{hBN/Si}}$ after several fitting methods comparison. Report $\kappa_{\text{i,hBN}}$ = 396 W m$^{-1}$ K$^{-1}$, $\kappa_{\text{o,hBN}}$ = 3.7 W m$^{-1}$ K$^{-1}$, and both $G$ values near $5.0 \times 10^7$ W m$^{-2}$ K$^{-1}$. |
| E-11 | Execute $f$-sweep fitting for Au/graphite using the given dataset. Targets: $\kappa_{\text{o,Gr}}$ and $G$. (One outlier data present, together with data problems in low-frequency ranges) | Use prompt-specified properties and spot size. Exclude one anomalous repeat. Report $\kappa_{\text{o,Gr}}$ = 16.1 W m$^{-1}$ K$^{-1}$ and $G = 3.8 \times 10^7$ W m$^{-2}$ K$^{-1}$. Identify and warn about the low-frequency data shift which gives a high fitting residual. |
| E-12 | Execute $f$-sweep fitting for Au/fused silica. Targets: $\kappa_{\text{Au}}$ and $G$. (Incorrect Au thickness provided) | Use the prompt-stated 50 nm Au thickness and report $\kappa_{\text{Au}}$ = 203 W m$^{-1}$ K$^{-1}$ and $G_{\text{Au/SiO}_2} = 9.5 \times 10^7$ W m$^{-2}$ K$^{-1}$. Identify the poor fit problem and try co-fitting the heat capacity of Au to reduce the residual. Warn that the fitted Au heat capacity deviates from the provided value. |

memory usage on the host remains low throughout, indicating that execution is not bottlenecked by the underlying hardware.

A run is deemed successful only when it satisfies all of the following criteria: completion within the time limit, generation of the required outputs, and fitted results falling within predefined numerical tolerances. Based on manual inspection of representative successful and failed execution traces, the numerical tolerances are set to 0.5% for most fitted quantities, which is sufficient to separate procedural failures from intrinsic numerical fitting errors, except for the spot size uncertainty where the tolerance is relaxed to 5% owing to its larger numerical error. For the L1 synthetic tasks, the ground truth used for the forward data generation serves as the reference, with the time limit set to 600 s; for the L2 real-data tasks, the target values are instead the analysis results obtained by human experts using the same FDTR backend, and the time limit is extended to 900 s. Given the stochastic nature of agent behavior, each task was independently repeated 10 times to sample as broad a range of outcomes as possible and to obtain statistically meaningful results. For each task, the success rate, token cost, and wall-clock time are recorded.

## 5. Results and discussion

### 5.1. Reliability in well-defined FDTR tasks

We first evaluate success rates on the L1 and L2 benchmark tests. Fig. 6 shows the number of successful runs out of 10 repetitions for each task, which reflects both numerical correctness and repeated-run reliability. The confidence intervals of the per-task success rates are provided in Fig. S1 of the Supplementary Information. Across both benchmark levels, Vibe-FDTR consistently achieves the highest success rates in all tasks, while the Code-agent and Agent-only references become progressively less reliable as task complexity increases, which is further supported by the distribution of the maximum relative fitting errors across all runs shown in Fig. S2a of the Supplementary Information.

For the L1 single-step tasks, Vibe-FDTR completes all 70 test runs successfully—a success rate of 100%, as shown in Fig. 6a. Code-agent achieves a success rate of 91.4%, showing that access to the domain code was sufficient for most explicitly specified fitting tasks. Its remaining failures are concentrated in sensitivity and uncertainty analyses, where correct execution requires more complicated configuration files and interface use. Agent-only achieves a success rate of only 38.6%, with successful runs limited mainly to simpler fitting tasks. These results indicate that the code layer of Vibe-FDTR provides the primary foundation for reliable single-step analysis, while structured procedural guidance further guarantees the success of tasks with more demanding requirements.

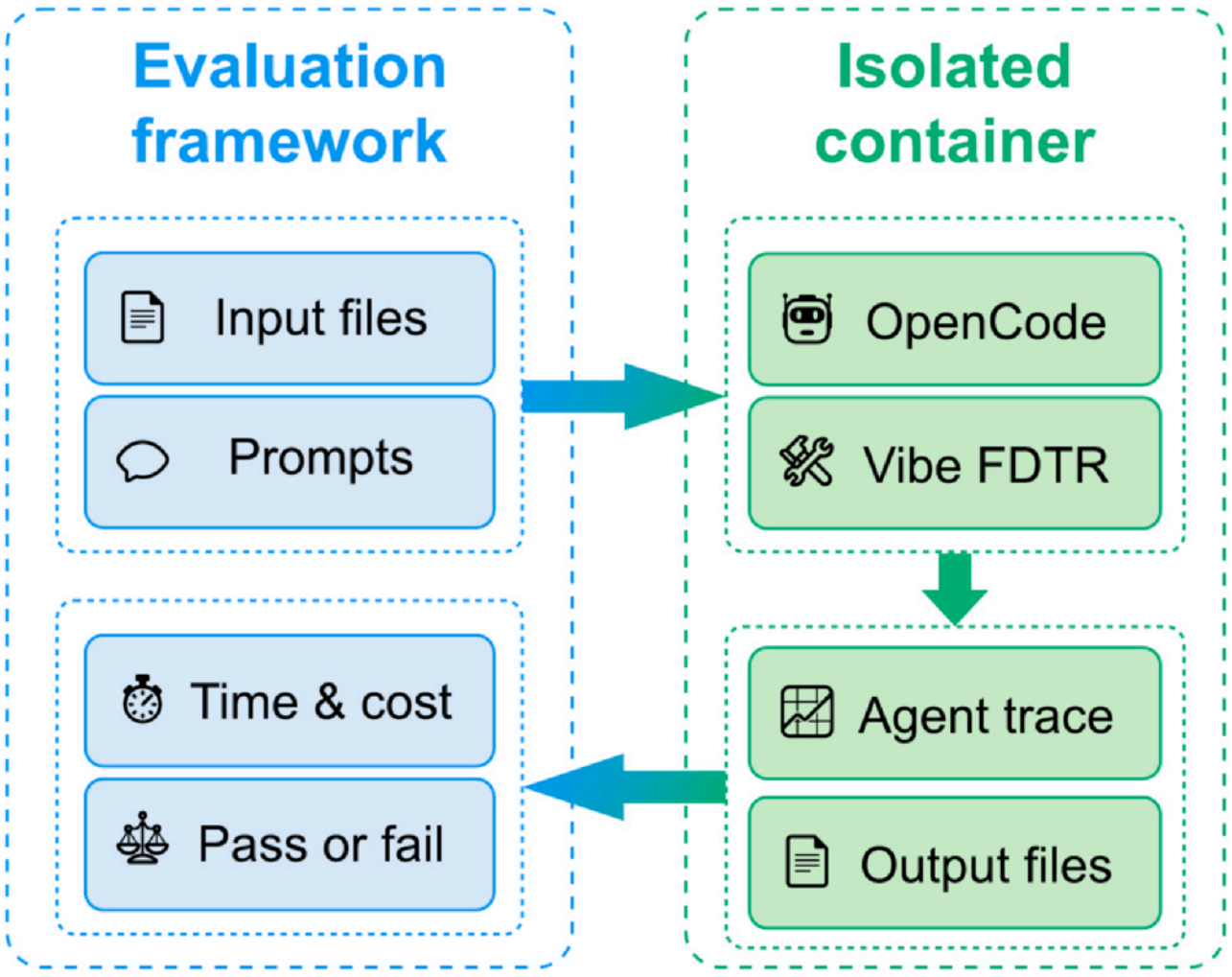

**Fig. 5.** Design of the evaluation environment for benchmark tasks.

The distinction becomes much more prominent for the L2 real-data workflows. Vibe-FDTR maintains a rather high success rate of 98.9%, with only one failed run out of 90, whereas Code-agent only succeeds in 36.7% of total runs and Agent-only produces no successful runs at all. Looking into the details of Fig. 6b, Code-agent performs comparatively better on the more direct data-processing and fitting tasks, but its reliability decreases sharply for composite, iterative, and batch workflows that require uncertainty propagation, parameter transfer, or multiple analysis stages. The contrast between L1 and L2 shows that the procedural guidance layer of Vibe-FDTR becomes increasingly important when a complete FDTR workflow must preserve consistency across data handling, configuration updates, and successive calculations. In addition, the above results also indicate that finishing the complex FDTR workflows from scratch remains challenging for current frontier open-source large language models. To examine the robustness of these conclusions against the choice of LLM, we further run the full benchmark suite on custom-deployed GLM-5.2 and repeat a subset of representative tasks using the faster DeepSeek-V4-Flash-0731; both results lead to the same conclusions as presented in Figure S3 and Table S1 of the Supplementary Information. Moreover, Vibe-FDTR with a locally deployed lightweight model, Qwen-3.8–27B, also delivers high reliability, as shown in Table S1 of the Supplementary Information, highlighting the auxiliary role of the framework for weaker models and unlocking the possibility of directly integrating LLMs into readily-accessible instrument control systems.

### 5.2. Failure mode analysis

To better understand the large performance gap between Vibe-FDTR and the two reference variants, we examined the failure modes in the L1 and L2 benchmarks. For L1 tasks, Code-agent failures arise mainly from errors in the configuration setup and command-line interface (CLI) usage, including bypassing the existing CLI tools and consequently misusing the underlying code functions in L1-F01, passing parameters in an invalid format in L1-S01, and mixing the parameter naming for isotropic and anisotropic materials in L1-U01. Although minor, these operational mistakes lead to wrong outputs and can largely be eliminated by the explicit configuration guidance as implemented in Vibe-FDTR. For the Agent-only configuration, the agent has to rely on the general knowledge of the model to interpret each task and develop dedicated code, leading to two main types of failure. First, substantial reasoning and execution resources are spent on model derivation and debugging. For example, in L1-U01, the agent spends much of its reasoning context to derive Jacobian matrices and uncertainty-propagation formulas, while in the L1-F03 and F04 tasks, repeated code modification and looping eventually trigger debugging timeouts. Second, the agent meets the challenges of concepts understanding, such as thermal conductivity and thermal resistance (L1-S01).

Moving to L2 tasks, we first look at the only failed run of Vibe-FDTR occurred in L2-C01, where the agent omits the requested frequency range, leading to a 1.2% deviation of the fitted result. This isolated configuration error does not indicate a systematic limitation. For Code-agent, the relatively low success rates already observed in the L1 sensitivity and uncertainty tasks carry over to L2 workflows containing these analysis steps. Further failures arise in iterative multi-step fitting and batch processing, where the agent often bypasses the intended configuration workflow, uses inappropriate initial values or bounds, or fails to maintain consistent parameter transfer across stages and temperatures. By contrast, Code-agent achieves relatively high success rates in L2-D01 and D02 tasks because they mainly extend the simple L1 fitting tasks with clearly specified data-selection and preprocessing steps, without introducing additional analysis dependencies. Agent-only completes no L2 task, as most runs exhaust their time budgets planning the multi-step tasks, exploring data directories, and deriving the underlying physical model before fitting could even begin.

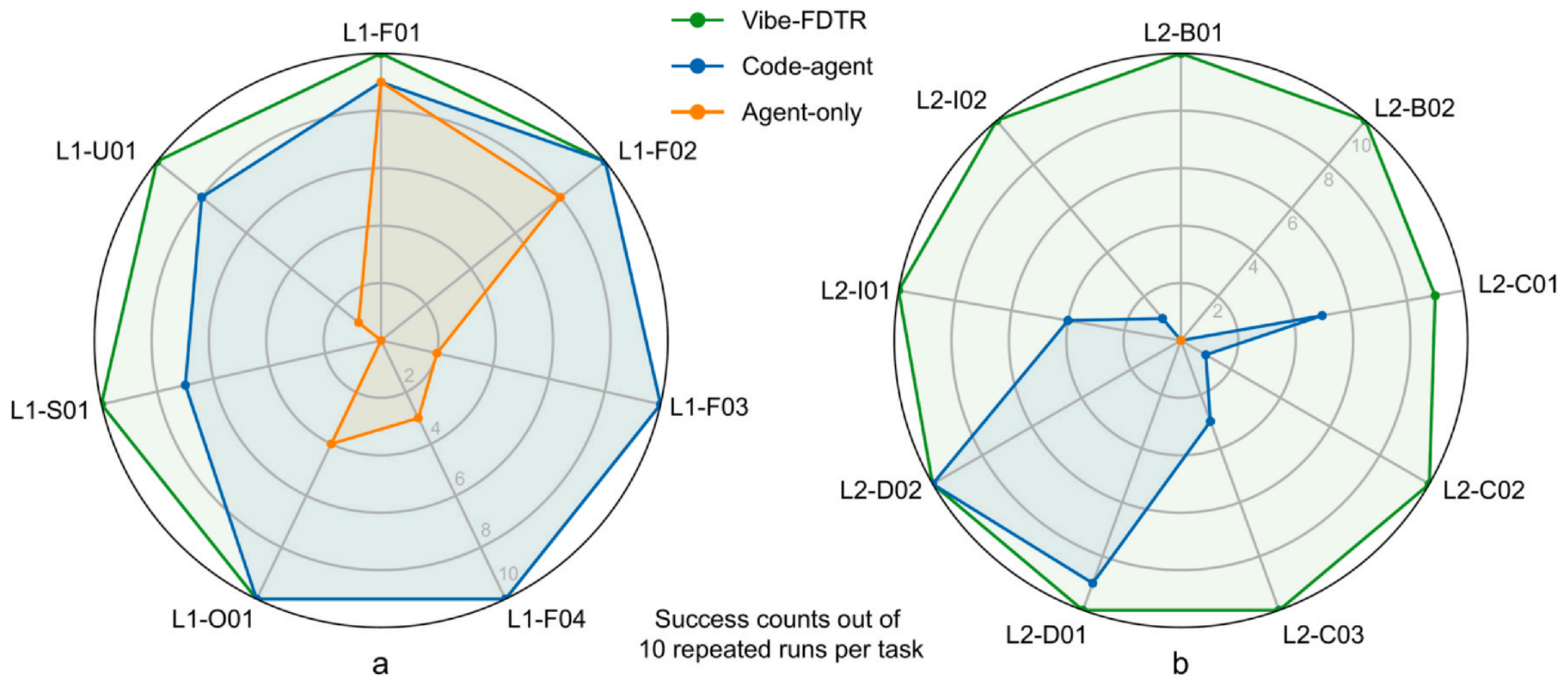

**Fig. 6.** Number of successful runs of L1 (a) and L2 (b) benchmark tasks for Vibe-FDTR and the two reference variants. Each task is repeatedly run for 10 times.

Reviewing these failure modes, the core value of Vibe-FDTR becomes evident: the framework encapsulates invaluable human-expert experience into skills together with the physical model and domain code, enabling domain-specific tasks to be executed efficiently and stably, and substantially saving the trial-and-error costs of writing relevant programs from scratch.

### 5.3. *Cost and efficiency*

We then analyze the token cost and time consumption during the L1 and L2 benchmark tasks. The distributions of the wall-clock time and per-run cost across all benchmark tasks are provided in Fig. S2(b) and Fig. S2c of the Supplementary Information. As shown in Fig. 7a, Vibe-FDTR requires only $0.0098 per L1 run and $0.0133 per L2 run, representing an 87.7% reduction relative to Code-agent at both benchmark levels. Agent-only also shows a lower cost than Code-agent, however, this should not be interpreted as greater efficiency since many of its runs simply fail or terminate before completing the full analysis. The much higher cost for Code-agent can be attributed mainly to the parallel inspection of the source code and repeated exploration of available interfaces, which substantially increase token consumption.

In Fig. 7b, we further demonstrate the average time consumption in the benchmark tests. Vibe-FDTR completes the L1 and L2 tasks with average runtimes of 88 and 106 s, respectively. These values are 64.9% and 69.0% lower than those for Code-agent, and 78.1% and 85.8% lower than those for Agent-only. In contrast to the cost, time consumption of Code-agent is lower than Agent-only due to the parallel code inspection process. Agent-only typically follows a more sequential process of constructing, testing, and debugging the code and the analysis workflow, resulting in the longest execution time. Because unsuccessful runs are stopped at preset time limits, the reported Agent-only runtime provides a conservative estimate of this overhead.

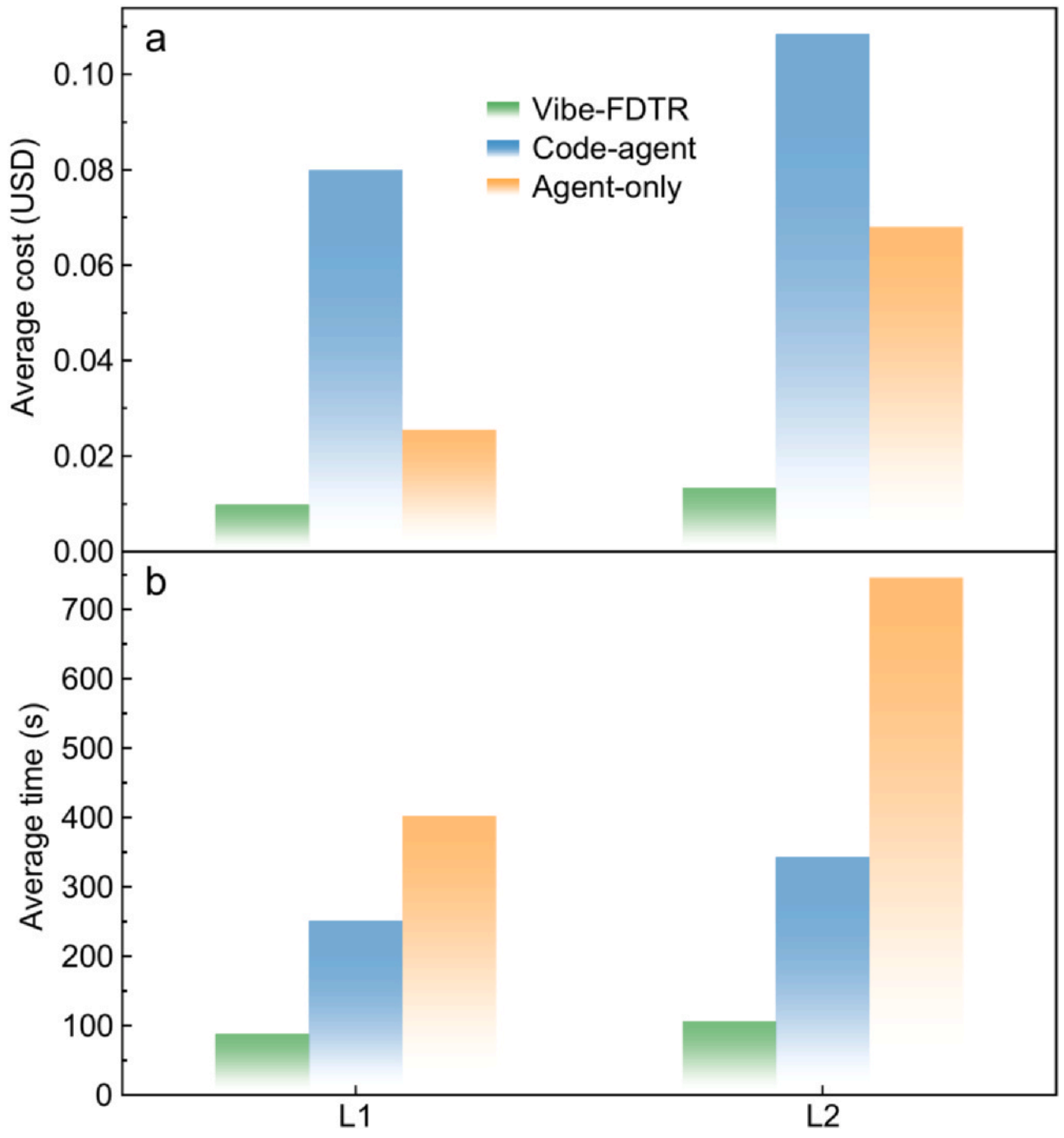

**Fig. 7.** Average cost (a) and time (b) per run for L1 and L2 benchmark tasks using Vibe-FDTR and the two reference variants.

To gain an intuitive understanding of the efficiency advantage of Vibe-FDTR, we take L2-B02, the most complex batch-processing task as an example. Vibe-FDTR completes this task within 2 min on average, which involves a full iterative fitting workflow at room temperature (RT, 295 K), then extending the fitting to eight additional temperature datasets and illustrating the temperature-dependent thermal conductivity as shown in the inset of Fig. 1. Even with ready-made analysis codes, the same sequence would normally require repeated manual efforts and much longer hands-on processing time.

The above results clearly demonstrate the advantages of Vibe-FDTR not only in its reliability but also in efficiency and cost savings, which holds great potential as a copilot for well-defined FDTR tasks.

### 5.4. *Expert mode demonstration*

Unlike the L1 and L2 benchmarks, the optional expert mode is evaluated qualitatively with a focus on scientifically reasonable decision process and final recommendation. As shown in Table 2, the E tasks involve eight design-only and four data-backed cases, covering a wide range of situations and challenges encountered by researchers in practice. Generally, Vibe-FDTR in the expert mode transforms open-ended user intent into explicit analysis assumptions and executable evaluation schemes. The recommendations are broadly consistent with expert experiences. In particular, Vibe-FDTR demonstrates its potential as an intelligent copilot in recommending complementary measurement modes, identifying poorly separable parameters, and recognizing abnormal data or unreliable fitting conditions.

To gain further insights, we look into the execution trace of the agent. Here, the E-7 task is used as an example and the main agent decisions and outputs are detailed in Fig. 8. First, the agent frames the task as a design-only problem and addresses the missing information of the thin film by adopting the properties of Si, while assuming a fixed thermal boundary conductance for $G_1$. Next, the agent organizes the design space according to the competing serial thermal resistances, $d_2/\kappa_2$ and $1/G_3$, and performs batch sensitivity calculations over broad ranges of film thickness and thermal conductivity. The agent discovers that the sensitivity curves of $\kappa_2$ and $G_3$ closely follow each other, and further validates the inseparability of these two parameters via representative uncertainty analysis. Moreover, the agent also tries alternative fitting schemes, including beam-offset fitting and separate single-parameter fits before concluding that a single FDTR measurement could not reliably recover both quantities simultaneously. Finally, the agent gives a structured recommendation including clear statement on the assumptions and risks. This execution aligns closely with the intended expert-mode workflow in Fig. 4.

The execution traces nevertheless reveal limitations in the scientific decisions made during this process. In E-7, the original user prompt does not specify which interface should be independently constrained. The agent jumps to fix $G_1$ and focuses on the separability of $\kappa_2$ and $G_3$, which is counterintuitive for human experts, although the final recommendation acknowledges the risk of an uncalibrated $G_1$. Similarly, in several data-backed cases, the backtracking process is also incomplete. For instance, E-10 retains both interfacial thermal conductances in an already coupled anisotropic fitting; E-11 recognizes the low-frequency inconsistency but does not redefine the fitting window and repeat the analysis using the more reliable high-frequency range; and E-12 improves the fit by allowing the Au heat capacity to vary without systematically testing whether the supplied transducer thickness is correct. These cases show that the expert mode is not yet capable of fully substituting for human expertise. We also note that the performance of the expert mode strongly depends on the underlying large language model of the agent.

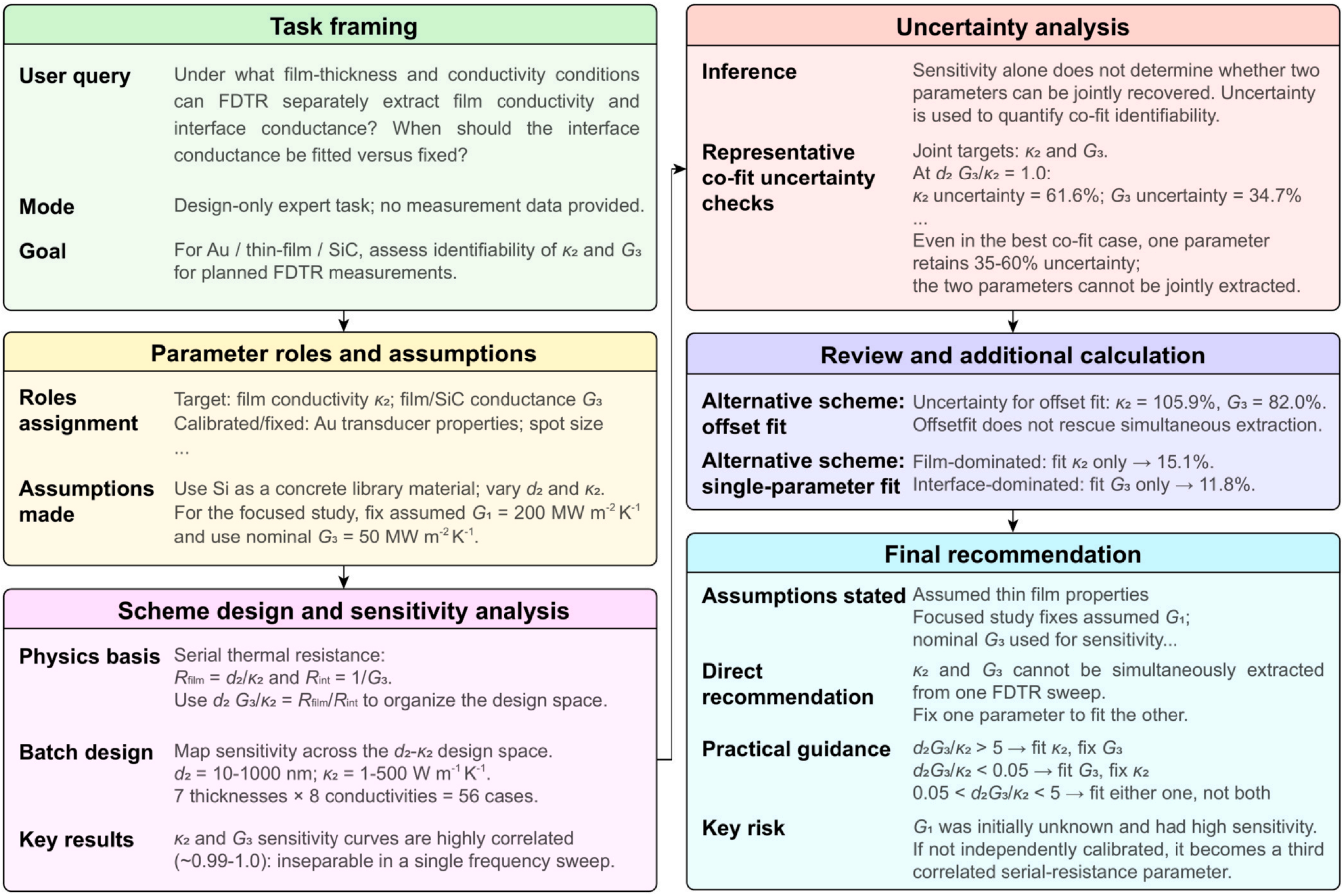

**Fig. 8.** Representative execution trace under the optional expert mode of Vibe-FDTR. The E-7 task is demonstrated as an example, and the agent outputs are selected and reorganized for clarity.

## 6. Conclusion

In summary, we have developed Vibe-FDTR, an agent-oriented framework that enables state-of-the-art LLM agents to perform reliable and reproducible FDTR analyses directly from natural language requests. To quantitatively assess its performance, we design a controlled two-level benchmark comprising single-step tasks (L1) with synthetic data and multi-step tasks (L2) based on real measurements of gold-coated graphite samples, together with a repeated-run evaluation protocol. Agents using Vibe-FDTR achieve success rates of 100% and 98.9% on the L1 and L2 tasks, respectively, compared to 91.4% and 36.7% for the Code-agent variant and 38.6% and 0% for the Agent-only setting. The per-run cost and runtime are also reduced by 87.7% and over 60%, respectively, relative to the Code-agent variant. These controlled comparisons show that the domain code package provides the physical foundation, while explicit procedural guidance via agent skills becomes increasingly important for maintaining reliability and efficiency as task complexity grows. Beyond executing well-specified instructions, the optional expert mode autonomously plans and performs sensitivity and uncertainty analyses and formulates physically grounded recommendations for underspecified tasks, although its scientific decisions still fall short of those of human experts and remain sensitive to the underlying language model. By encapsulating validated domain code and expert procedural knowledge into agent skills, Vibe-FDTR lowers the barrier to FDTR analysis for both new practitioners and specialists from other fields. This framework can be readily extended to other thermoreflectance techniques such as TDTR, and integrated with instrument control and data acquisition, offering a concrete route toward fully autonomous thermoreflectance metrology.

## Author contributions

Yang F and Song B conceived the idea and designed the research. Yang F developed the Vibe-FDTR package, designed the benchmark tasks, and performed the FDTR measurement on graphite. Li W improved the packages and designed the benchmark test environment. Yang F and Li W wrote the original draft of the manuscript. Song B supervised the work. All authors reviewed and edited the manuscript.

## CRediT authorship contribution statement

**Fuwei Yang:** Writing – review & editing, Writing – original draft, Software, Methodology, Conceptualization. **Weiheng Li:** Writing – review & editing, Writing – original draft, Validation, Software, Investigation, Data curation. **Bai Song:** Writing – review & editing, Supervision, Funding acquisition, Conceptualization.

## Data availability

The agent inputs, benchmark runner, and all execution traces of the runs presented in this work are deposited in the benchmark repository at https://github.com/lwh9346/Vibe_FDTR_benchmarks; the Vibe-FDTR source code and skills are publicly available at https://github.com/yfw-sunny/Vibe_FDTR. Both repositories are hosted on GitHub.

## Declaration of Competing Interest

The authors declare that they have no known competing financial interests or personal relationships that could have appeared to influence the work reported in this paper. Bai Song is an Editorial Board Member for Advanced Scientific Instruments and was not involved in the editorial review or the decision to publish this article.

## Declaration of Generative AI and AI-assisted technologies in the writing process

During the preparation of this work, the authors used OpenCode and DeepSeek API service in order to assist coding and language polishing, and used codex to generate some visual elements in the concept figures. After using this tool/service, the authors reviewed and edited the content as needed and take full responsibility for the content of the published article.

## Acknowledgments

This work was supported by the Science Fund for Creative Research Groups from the National Natural Science Foundation of China (No. 52521007), the Scientific Research Innovation Capability Support Project for Young Faculty (ZYGXQNJSKYCXNLZCXM-E1) from the Ministry of Education of China, the National Key R&D Project from the Ministry of Science and Technology of China (No. 2022YFA1203100 and No. 2024YFA1207900), and the High-performance Computing Platform of Peking University. We thank Shuangdui Wu for her help with the figures. B.S. acknowledges support from the New Cornerstone Science Foundation through the XPLORER PRIZE.

## Appendix A. Supporting information

Supplementary data associated with this article can be found in the online version at doi:10.1016/j.asi.2026.100049.